\documentclass[fleqn,12pt,twoside]{article}
\usepackage{espcrc1}
\usepackage{graphics}

\usepackage{graphicx}
\usepackage[figuresright]{rotating}

\newcommand{\AmS}{{\protect\the\textfont2
  A\kern-.1667em\lower.5ex\hbox{M}\kern-.125emS}}

\title{\bf Models for RHIC and LHC: New Developments}

\author{K. Werner \address[SUB]{SUBATECH, Universit\'e de
Nantes -- IN2P3/CNRS -- EMN,  Nantes, France},
H. J. Drescher \address[NY]{Physics Department, New York University, 
New York, USA},
S. Ostapchenko \address[FZ]{Inst. f. Kernphysik, Forschungszentrum Karlsruhe, 
Karlsruhe, Germany},
and T. Pierog\addressmark[SUB]
}

\newcommand{\LyX}{L\kern-.1667em\lower.25em\hbox{Y}\kern-.125emX\@}

\begin{document}

\maketitle

\begin{abstract}
We outline inconsistencies in presently used models for high energy nuclear 
scattering, which make their application quite unreliable. Many "successes" are 
essentially based on an artificial freedom of parameters, which does not exist
when the models are constructed properly.

The problem is the fact that any multiple scattering theory
requires an appropriate treatment of the energy sharing between the individual
interactions, which is technically very difficult to implement.
Lacking a satisfying solution to this problem, it has been simply ignored.
 
We introduce a fully self-consistent formulation of the multiple-scattering
scheme. Inclusion of soft and hard 
components -- very crucial at high energies -- appears in a "natural way", 
providing a smooth transition from soft to hard physics.

We can show that the effect of appropriately considering energy 
conservation has a big influence on the results, and MUST therefore be included 
in any serious calculation.

\end{abstract}

\section{Open Problems}

With the start of the RHIC program to investigate nucleus-nucleus collisions
at very high energies, there is an increasing need of computational tools in
order to provide a clear interpretation of the data. The situation is not satisfactory
in the sense that there exists a nice theory (QCD) but we are not able to treat
nuclear collisions strictly within this framework, and on the other hand there
are simple models, which can be applied easily but which have no solid theoretical
basis. A good compromise is provided by effective theories, which are not derived
from first principles, but which are nevertheless self-consistent and calculable.
A candidate seems to be the Gribov-Regge approach, and -- being formally quite
similar -- the eikonalized parton model. Here, however, some inconsistencies
occur, which we are going to discuss in the following, before we provide a solution
to the problem.

Gribov-Regge theory \cite{gri68,gri69} is by construction a multiple scattering
theory. The elementary interactions are realized by complex objects called ``Pomerons'',
who's precise nature is not known, and which are therefore simply parameterized,
with a couple of parameters to be determined by experiment \cite{dre00}. Even
in hadron-hadron scattering, several of these Pomerons are exchanged in parallel
(the cross section for exchanging a given number of Pomerons is called''topological
cross section''). Simple formulas can be derived for the (topological) cross
sections, expressed in terms of the Pomeron parameters.

In order to calculate exclusive particle production, one needs to know how to
share the energy between the individual elementary interactions in case of multiple
scattering. We do not want to discuss the different recipes used to do the energy
sharing (in particular in Monte Carlo applications). The point is, whatever
procedure is used, this is not taken into account in the calculation of cross
sections discussed above \cite{bra90},\cite{abr92}. So, actually, one is using
two different models for cross section calculations and for treating particle
production. Taking energy conservation into account in exactly the same way
will modify the (topological) cross section results considerably. 

Being another popular approach, the parton model \cite{sjo87} amounts to presenting
the partons of projectile and target by momentum distribution functions, \( f_{i} \)
and \( f_{j} \), and calculating inclusive cross sections for the production
of parton jets as a convolution of these distribution functions with the elementary
parton-parton cross section \( d\hat{\sigma }_{ij}/dp_{\perp }^{2} \), where
\( i,j \) represent parton flavors. This simple factorization formula is the
result of cancelations of complicated diagrams and hides therefore the complicated
multiple scattering structure of the reaction, which is finally recovered via
some unitarization procedure. The latter one makes the approach formally equivalent
to the Gribov-Regge one and one therefore encounters the same conceptual problems
(see above).

\section{A Solution: Parton-based Gribov-Regge Theory}

As a solution of the above-mentioned problems, we present a new approach which
we call ``Parton-based Gribov-Regge Theory'': we have a consistent
treatment for calculating (topological) cross sections and particle production
considering energy conservation in both cases; in addition, we introduce hard
processes in a natural way. 

The basic guideline of our approach is theoretical consistency. We cannot derive
everything from first principles, but we use rigorously the language of field
theory to make sure not to violate basic laws of physics, which is easily done
in more phenomenological treatments (see discussion above).

Let us consider nucleus-nucleus (\( AB \)) scattering. In the Glauber-Gribov
approach \cite{gla59,gri69}, the nucleus-nucleus scattering amplitude is defined
by the sum of contributions of diagrams, corresponding to multiple elementary
scattering processes between parton constituents of projectile and target nucleons.
These elementary scatterings are the sum of
soft, semi-hard, and hard contributions: \( T_{2\rightarrow 2}=T_{\mathrm{soft}}+T_{\mathrm{semi}}+T_{\mathrm{hard}} \).
A corresponding relation holds for the inelastic amplitude \( T_{2\rightarrow X} \).
A cut elementary diagram will be graphically
represented by a vertical dashed line, whereas the elastic amplitude 
by an unbroken line:

\vspace{0.3cm}
{\par\centering \resizebox*{!}{0.05\textheight}{\includegraphics{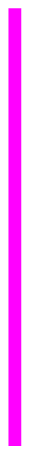}} \( \quad  \)\( =T_{2\rightarrow 2} \),
\( \quad  \)\( \quad  \)\resizebox*{!}{0.05\textheight}{\includegraphics{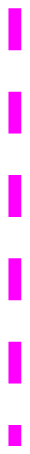}} \( \quad  \)\( =\sum _{X}(T_{2\rightarrow X}) \)\( (T_{2\rightarrow X})^{*} \).\par}
\vspace{0.3cm}

\begin{figure}[htb]
{\par\centering \resizebox*{!}{0.24\textheight}{\includegraphics{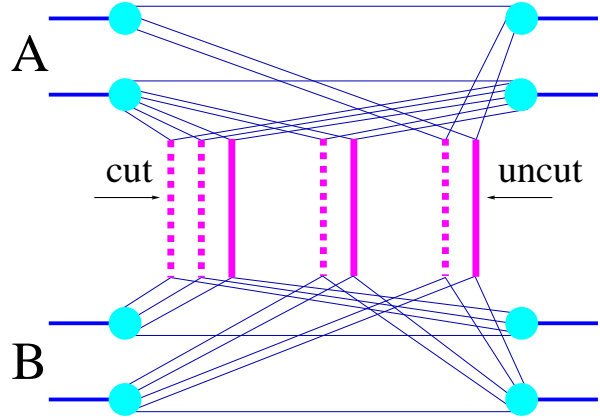}} \par}

\caption{Typical interference term contributing to the squared inelastic amplitude.\label{grtppaac}}
\end{figure}

\noindent This is very handy for treating the nuclear scattering model. We define
the model via the elastic scattering amplitude \( T_{AB\rightarrow AB} \) which
is assumed to consist of purely parallel elementary interactions between partonic
constituents. The amplitude is therefore a sum of such terms.
One has to be careful about energy conservation: all the partonic constituents
leaving a nucleon have to share the momentum of the nucleon.
So in the explicit formula one has an integration over momentum fractions of
the partons, taking care of momentum conservation. Having defined elastic scattering,
inelastic scattering and particle production is practically given, if one employs
a quantum mechanically self-consistent picture. Let us now consider inelastic
scattering: one has of course the same parallel structure, just some of the
elementary interactions may be inelastic, some elastic. The inelastic amplitude
being a sum over many terms -- \( T_{AB\rightarrow X}=\sum _{i}T^{(i)}_{AB\rightarrow X} \)
-- has to be squared and summed over final states in order to get the inelastic
cross section, which provides interference terms \( \sum _{X}(T^{(i)}_{AB\rightarrow X})(T^{(j)}_{AB\rightarrow X})^{*} \),
which can be conveniently expressed in terms of the cut and uncut elementary
diagrams, as shown in fig. \ref{grtppaac}. So we are doing nothing more than
following basic rules of quantum mechanics.
Of course a diagram with 3 inelastic elementary interactions does not interfere
with the one with 300, because the final states are different. So it makes sense
to define classes \( K \) of interference terms (cut diagrams) contributing
to the same final state, as all diagrams with a certain number of inelastic
interactions and with fixed momentum fractions of the corresponding partonic
constituents. One then sums over all terms within each class \( K \), and obtains
for the inelastic cross section

\noindent 
\[
\sigma _{AB}(s)=\int d^{2}b\: \sum _{K}\Omega ^{(s,b)}(K)\]
 where we use the symbolic notation \( d^{2}b=\int d^{2}b_{0}\, \int d^{2A}b_{A}\, \rho (b_{A})\, \int d^{2B}b_{B}\, \rho (b_{B}) \)
which means integration over impact parameter \( b_{0} \) and in addition averaging
over nuclear coordinates for projectile and target. The variable \( K \) is
characterized by \( AB \) numbers \( m_{k} \) representing the number of cut
elementary diagrams for each possible pair of nucleons and all the momentum
fractions \( x^{+} \) and \( x^{-} \) of all these elementary interactions
(so \( K \) is a partly discrete and partly continuous variable, and \( \sum  \)
is meant to represent \( \sum \int  \)). This is the really new and very important
feature of our approach: we keep explicitly the dependence on the longitudinal
momenta, assuring energy conservation at any level of our calculation. 

The calculation of \( \Omega  \) actually very difficult and technical, but
it can be done and we refer the interested reader to the literature \cite{dre00}. 

The quantity \( \Omega ^{(s,b)}(K) \) can now be interpreted as the probability
to produce a configuration \( K \) at given \( s \) and \( b \). So we have
a solid basis for applying Monte Carlo techniques: one generates configurations
\( K \) according to the probability distribution \( \Omega  \) and one may
then calculate mean values of observables by averaging Monte Carlo samples.
The problem is that \( \Omega  \) represents a very high dimensional probability
distribution, and it is not obvious how to deal with it. We decided to develop
powerful Markov chain techniques \cite{hla98} in order to avoid to introduce
additional approximations.

\section{Summary}

We provide a new formulation of the multiple scattering mechanism in nucleus-nucleus
scattering, where the basic guideline is theoretical consistency. We avoid in
this way many of the problems encountered in present day models. We also introduce
the necessary numerical techniques to apply the formalism in order to perform
practical calculations.

This work has been funded in part by the IN2P3/CNRS (PICS 580) and the Russian
Foundation of Basic Researches (RFBR-98-02-22024). 

\bibliographystyle{pr2}
\bibliography{a}

\begin{thebibliography}{1}

\bibitem{gri68}
V.~N. Gribov,
\newblock Sov. Phys. JETP {\bf 26}, 414 (1968).

\bibitem{gri69}
V.~N. Gribov,
\newblock Sov. Phys. JETP {\bf 29}, 483 (1969).

\bibitem{dre00}
H.~J. Drescher, M.~Hladik, S.~Ostapchenko, T.~Pierog, and K.~Werner,
\newblock (2000), hep-ph/0007198,
\newblock to be published in Physics Reports.

\bibitem{bra90}
M.~Braun,
\newblock Yad. Fiz. (Rus) {\bf 52}, 257 (1990).

\bibitem{abr92}
V.~A. Abramovskii and G.~G. Leptoukh,
\newblock Sov.J.Nucl.Phys. {\bf 55}, 903 (1992).

\bibitem{sjo87}
T.~Sjostrand and M.~van Zijl,
\newblock Phys. Rev. {\bf D36}, 2019 (1987).

\bibitem{gla59}
R.~J. Glauber,
\newblock {\em in Lectures on theoretical physics} (N.Y.: Inter-science
  Publishers, 1959).

\bibitem{hla98}
M.~Hladik,
\newblock {\em Nouvelle approche pour la diffusion multiple dans les
  interactions noyau-noyau aux \'energies ultra-relativistes},
\newblock PhD thesis, Universit\'e de Nantes, 1998.

\end{thebibliography}

\end{document}